\documentclass[apjl]{emulateapj}

\def\kms{\ifmmode{\rm km\thinspace s^{-1}}\else km\thinspace s$^{-1}$\fi}
\def\ms{\ifmmode{\rm m\thinspace s^{-1}}\else m\thinspace s$^{-1}$\fi}

\shortauthors{Torres}
\shorttitle{GJ 436}

\begin{document}

\journalinfo{Accepted for publication in The Astrophysical Journal Letters}

\title{The transiting exoplanet host star GJ 436: a test of stellar
evolution models in the lower main sequence, and revised planetary
parameters}

\author{Guillermo Torres}

\affil{Harvard-Smithsonian Center for Astrophysics, 60 Garden St.,
Cambridge, MA 02138}

\email{gtorres@cfa.harvard.edu}

\begin{abstract} 

Knowledge of the stellar parameters for the parent stars of transiting
exoplanets is pre-requisite for establishing the planet properties
themselves, and often relies on stellar evolution models. GJ~436,
which is orbited by a transiting Neptune-mass object, presents a
difficult case because it is an M dwarf. Stellar models in this mass
regime are not as reliable as for higher mass stars, and tend to
underestimate the radius. Here we use constraints from published
transit light curve solutions for GJ~436 along with other
spectroscopic quantities to show how the models can still be used to
infer the mass and radius accurately, and at the same time allow the
radius discrepancy to be estimated. Similar systems should be found
during the upcoming \emph{Kepler\/} mission, and could provide in this
way valuable constraints to stellar evolution models in the lower main
sequence. The stellar mass and radius of GJ~436 are $M_{\star} =
0.452_{-0.012}^{+0.014}$~M$_{\sun}$ and $R_{\star} =
0.464_{-0.011}^{+0.009}$~R$_{\sun}$, and the radius is 10\% larger
than predicted by the standard models, in agreement with previous
results from well studied double-lined eclipsing binaries.  We obtain
an improved planet mass and radius of $M_p = 23.17 \pm
0.79$~M$_{\earth}$ and $R_p = 4.22_{-0.10}^{+0.09}$~R$_{\earth}$, a
density of $\rho_p = 1.69_{-0.12}^{+0.14}$ g~cm$^{-3}$, and an orbital
semimajor axis of $a = 0.02872 \pm 0.00027$ AU.
		
\end{abstract}

\keywords{
planetary systems ---
stars: evolution ---
stars: fundamental parameters ---
stars: individual (GJ~436) ---
stars: low-mass, brown dwarfs
}

\section{Introduction}
\label{sec:introduction}

The applications of stellar evolution theory to astrophysics are so
widespread, and its validity so often taken for granted, that it is
easy to forget that it took decades to develop, and significant effort
to validate by comparison with careful measurements, a process that
still continues. It is usually only when those theoretical predictions
fail that the classical discipline of stellar evolution ``makes the
headlines'', and even then it draws the attention of relatively
few. One such instance has occurred for low mass main-sequence
stars. Over the last 10 years or so it has become clear that our
understanding of the structure and evolution of these objects is still
incomplete.  Discrepancies between theory and observation in the radii
of stars under 1~M$_{\sun}$, first mentioned by \cite{Hoxie:73},
\cite{Lacy:77}, and others, are now well documented for several
low-mass eclipsing binaries \citep[see, e.g.,][]{Popper:97,
Clausen:99, Torres:02, Ribas:03, Lopez-Morales:05}. Differences in the
effective temperatures have been observed as well. The direction of
these disagreements is such that model radii are underestimated by
roughly 10\%, while effective temperatures are overestimated.

In recent years stellar evolution has had important applications in
the field of transiting extrasolar planets. This is because the
planetary parameters of interest (mass $M_p$, radius $R_p$) depend
rather directly on those of the star ($M_{\star}$, $R_{\star}$), and
in most cases models provide the only means of determining the
latter. The subject of this paper is GJ~436, a late-type star found by
\cite{Butler:04} to be orbited by a Neptune-mass planet with a period
of 2.644 days. This object was later discovered by \cite{Gillon:07a}
to undergo transits, enabling its size to be determined
($\sim$4~R$_{\earth}$).  As the only M dwarf among the 22 currently
known transiting planet host stars, GJ~436 (M2.5V) presents a special
challenge for establishing the stellar parameters, because of the
disagreements noted above. Little mention seems to have been made of
this, and for the most part past studies have relied instead on
empirical mass-luminosity ($M\!-\!L$) relations to set the mass of
GJ~436. Radius estimates have often rested on the assumption of
numerical equality between $M_{\star}$ and $R_{\star}$ for M
stars. Despite being the closest transiting planet system (only 10 pc
away), it is rather surprising that the mass of the star is only known
to about 10\% \citep{Maness:07, Gillon:07b}. This is currently
limiting the precision of the planetary mass, and some of that
uncertainty translates also to the radius. These two properties are
critical for studying the structure of the object.

Given the importance of GJ~436 as the parent star of the only
Neptune-mass transiting exoplanet found so far, and hence the closest
analog to our Earth with a mass and radius determination, one of the
motivations of this paper is to improve the precision of the stellar
and planetary parameters by making use of additional observational
constraints not used before. Specifically, we incorporate the
information on the stellar \emph{density} directly available from the
transit light curve \citep{Sozzetti:07}, which provides a strong
handle on the size of the star.

The nature of the discrepancies between evolutionary models and
observations for low-mass stars has been examined recently from both
the observational and theoretical points of view by
\cite{Lopez-Morales:07} and \cite{Chabrier:07}. Further progress
depends on gathering more evidence to supplement the few available
highly accurate mass and radius measurements based on double-lined
eclipsing binaries.  Thus, a second motivation for this work, despite
the fact that GJ~436 is not a double-lined eclipsing binary, is to
present a way of using all observational constraints simultaneously to
show that the star presents the same radius anomaly found for the
other systems, or more generally, to test the models.  Because similar
constraints may become available in the future for other M dwarfs
given the keen interest in finding smaller and smaller transiting
planets, we anticipate that the indirect technique described here may
yield valuable information on this problem and eventually help improve
our understanding of low-mass stars.

\section{Constraints and methodology}
\label{sec:method}

The basic procedure for establishing the mass and radius of a star
that is not in a double-lined eclipsing binary is to place it on an
H-R diagram using observational constraints, and compare it with
stellar evolution models. We adopt here the models by
\cite{Baraffe:98} for a mixing-length parameter of $\alpha_{\rm ML} =
1.0$, which are widely used for low-mass stars.  The constraints
available for GJ~436 are several. The spectroscopic study by
\cite{Maness:07} established the effective temperature to be $T_{\rm
eff} = 3350 \pm 300$~K, and the metallicity was estimated
photometrically by \cite{Bonfils:05} to be [Fe/H] $= -0.03 \pm
0.20$. The well-determined {\it Hipparcos} parallax is $\pi_{\rm HIP}
= 97.73 \pm 2.27$ mas. With available visual and near infrared
photometry from \cite{Leggett:88} and 2MASS, the absolute magnitudes
become $M_V = 10.610 \pm 0.051$ and $M_K = 6.048 \pm
0.052$. Additionally, in view of the difficulties in determining
effective temperatures for M dwarfs \citep[see, e.g.,][]{Maness:07},
we consider as well the infrared color $J\!-\!K = 0.802 \pm
0.024$. Following \cite{Carpenter:01} the 2MASS magnitudes have been
transformed to the CIT system of \cite{Elias:82} adopted in the
\cite{Baraffe:98} models, and averaged with those of
\cite{Leggett:88}, already on that system. The agreement between the
two sources is excellent.

Transit light curves for GJ~436 have been obtained from the ground by
\cite{Gillon:07a} in the $V$ band, and also at 8$\mu$m by
\cite{Deming:07} and \cite{Gillon:07b} using the Spitzer Space
Telescope. Aside from minor corrections due to limb-darkening, transit
light curves can in general be described using three parameters: the
radius ratio between the planet and the star ($R_p/R_{\star}$), the
normalized planet-star separation ($a/R_{\star}$), and the impact
parameter ($b \equiv a\cos i/R_{\star}$), where $i$ is the inclination
angle of the orbit. \cite{Seager:03} have shown that $a/R_{\star}$ is
directly related to the density of the star, and thus contains
valuable information on its size. \cite{Sozzetti:07} have described
how $a/R_{\star}$ can be used together with $T_{\rm eff}$ and stellar
evolution models to infer $M_{\star}$ and $R_{\star}$. Briefly, the
measured values of $a/R_{\star}$ and $T_{\rm eff}$ are compared with a
fine grid of model isochrones for a wide range of ages and
metallicities. Theoretical stellar properties are interpolated along
each isochrone using a small step in mass, and all points in the H-R
diagram matching the observations within their uncertainties are
recorded. The best-fitting mass and corresponding radius are assigned
errors based on the full range of model values that are consistent
with the observations. Other stellar properties can then be read off
the best fitting model. This is the procedure we apply below.  For
GJ~436 we have restricted the comparison to solar metallicity, given
the [Fe/H] estimate by \cite{Bonfils:05}.  The values of the light
curve parameters we adopt are weighted averages of the results from
the ground-based and Spitzer photometry: $a/R_{\star} = 13.34 \pm
0.58$, $R_p/R_{\star} = 0.0834 \pm 0.0007$, and $b = 0.848 \pm 0.010$.

\section{Mass and radius determinations}
\label{sec:mass}

In previous studies the mass and radius of GJ~436 have been derived in
three ways: either 1) the mass has been obtained from the
near-infrared ($JHK$) mass-luminosity relations of \cite{Delfosse:00}
\citep[giving $M_{\star} = 0.44 \pm 0.04$~M$_{\sun}$;][]{Maness:07}
and the radius has been assumed to be numerically equal to the mass
\citep{Gillon:07a}, or 2) the mass has been held fixed at the above
value and the radius constrained directly from the light curve
\citep{Gillon:07a, Gillon:07b}, or 3) both $M_{\star}$ and $R_{\star}$
have been solved for simultaneously subject to the constraint that
they be numerically equal and making use of the implicit sensitivity
$R \propto M^{1/3}$ between mass and radius in the light curve fitting
procedure \citep{Deming:07, Gillon:07b}. In the second case the radius
obtained is near 0.46~R$_{\sun}$ for $M_{\star} = 0.44$~M$_{\sun}$,
and in the third case $M_{\star} = R_{\star} \approx 0.47$ or 0.48 in
solar units. Thus, differences of order 0.02 to 0.04 remain between
these determinations, depending on the procedure used.

\begin{figure}
\vskip -0.3in 
\epsscale{1.15} 
\plotone{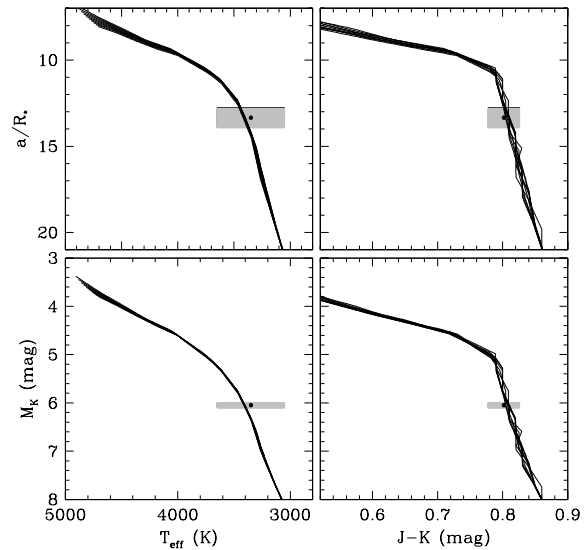}
\vskip -0.1in 

\figcaption[]{Observational constraints for GJ~436 provided by
$a/R_{\star}$, $T_{\rm eff}$, $M_K$, and $J\!-\!K$, shown in four
different combinations against \cite{Baraffe:98} isochrones for solar
composition. Ages between 1 and 10 Gyr are represented, although the
curves are hardly distinguishable because stars in this mass range
evolve very slowly.  The raggedness of the isochrones in the right
panels is due to the limited precision with which magnitudes are
tabulated in the published models. The agreement between the latter
and the observations appears good in all panels, but the implied
masses and radii show large systematic differences indicative of
internal inconsistencies in the models.
\label{fig:baraffe}}

\end{figure}

The methodology in the present paper is completely different, and can
yield improved precision and also give a better understanding of
possible systematics. We initially applied the $a/R_{\star}$ and
$T_{\rm eff}$ constraints as described in \S\ref{sec:method} using the
\cite{Baraffe:98} models, and obtained a mass value near $M_{\star} =
0.50$~M$_{\sun}$, which is considerably larger than previous
estimates. The predicted absolute visual magnitude ($M_V = 9.86$) is
also much brighter than that computed directly from the {\it
Hipparcos\/} parallax. This inconsistency strongly suggests a problem
with the models, which is not entirely unexpected for a star of this
type.  As an alternative to $a/R_{\star}$, we then experimented using
the absolute $K$ magnitude as a proxy for luminosity, as well as
replacing $T_{\rm eff}$ with the color index $J\!-\!K$.\footnote{We
have refrained from directly applying any constraint based on the $V$
magnitude because of suspected deficiencies in the models for optical
passbands, related to missing molecular opacity sources shortward of
1~$\mu$m \citep[see, e.g.,][]{Baraffe:98, Delfosse:00}. Experiments
using the $I\!-\!K$ color as an alternative indicate that the $I$ band
is also affected at some level.}  Figure~\ref{fig:baraffe} displays
the four constraints in different combinations against
solar-metallicity isochrones from 1 to 10 Gyr.  The seemingly good fit
in all planes belies the serious discrepancies present in other
derived quantities that are not shown explicitly. Those results are
listed in Table~\ref{tab:baraffe}. Masses inferred from $a/R_{\star}$
are systematically larger than those from $M_K$, and so are the
radii. The masses from $M_K$ come close to the estimates from the
empirical $M\!-\!L$ relations, but the corresponding radii are
considerably smaller than expected. This would seem to go in the
direction of the results from eclipsing binaries (see
\S\ref{sec:introduction}).

\begin{figure}
\vskip -0.3in 
\epsscale{1.15} 
\plotone{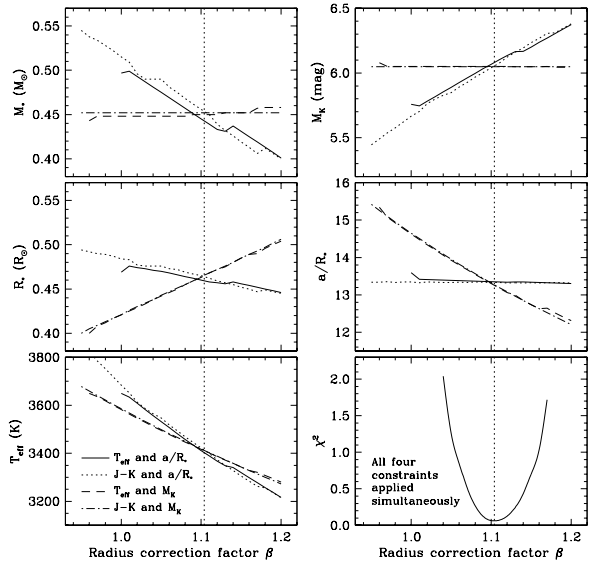}
\vskip 0.1in 

\figcaption[]{Stellar properties for GJ~436 derived from the
comparison with stellar evolution models by \cite{Baraffe:98}, as a
function of the adjustment factor applied to the model radii
($\beta$). The observational constraints given by $a/R_{\star}$,
$T_{\rm eff}$, $M_K$, and $J\!-\!K$ are applied and shown in pairwise
combinations, as labeled. Kinks in the curves are a reflection of the
discreteness of some of the quantities tabulated in the models. The
simultaneous application of all four constraints results in the
$\chi^2$ curve shown at the bottom right, indicating a best fit for a
$\beta$ value near 1.1 (see text). This value is represented with the
vertical dotted line running through this and the other panels.
\label{fig:beta}}

\end{figure}

On the other hand, there is good evidence from various sources that
the bolometric luminosities from these models are not seriously in
error \citep{Delfosse:00, Torres:02, Ribas:06, Torres:06}.  This
suggests that adjustments to the model radii and temperatures might
resolve the discrepancies in Table~\ref{tab:baraffe}, and allow us to
obtain a meaningful result for GJ~436. We explored this by introducing
a correction factor $\beta$ to the radii, and at the same time
applying a factor $\beta^{-1/2}$ to the temperatures in order to
preserve the bolometric luminosity. We repeated the comparison between
the adjusted models and each of the four sets of constraints for a
range of $\beta$ factors centered on the value indicated by the
eclipsing binary studies.  Figure~\ref{fig:beta} shows the result for
several of the key stellar properties. The lines corresponding to the
four sets of constraints seem to converge for a value of $\beta$ near
1.1 (representing a 10\% correction to the model radii), which happens
to be the typical factor found by the eclipsing binary studies
mentioned earlier. For this value of $\beta$ the models yield
essentially the same mass, radius, and luminosity for GJ~436,
independently of which set of observational constraints is used, as
one would expect from a realistic model. Thus, a self-consistent
solution is achieved. To arrive at the best possible values of
$M_{\star}$ and $R_{\star}$ we next applied all the constraints
simultaneously, and varied $\beta$ as before, seeking the best
agreement with the measurements. The result is illustrated in the
bottom right panel of Figure~\ref{fig:beta}, where the quality of the
match as represented by $\chi^2$ is shown as a function of the
correction factor over the restricted range in which the models agree
with all four observables within their errors. The best match is again
near $\beta = 1.1$. The resulting mass and radius are $M_{\star} =
0.452_{-0.012}^{+0.014}$~M$_{\sun}$ and $R_{\star} =
0.464_{-0.011}^{+0.009}$~R$_{\sun}$. These and other inferred stellar
properties for GJ~436 are listed in the top section of
Table~\ref{tab:properties}. We emphasize that these quantities are the
result of the simultaneous application of the four constraints, and
the agreement with some of the values in Table~\ref{tab:baraffe} is
accidental.

\section{The radius disagreement with the models}
\label{sec:disagreement}

\begin{figure}
\vskip -0.4in 
\epsscale{1.25} 
\plotone{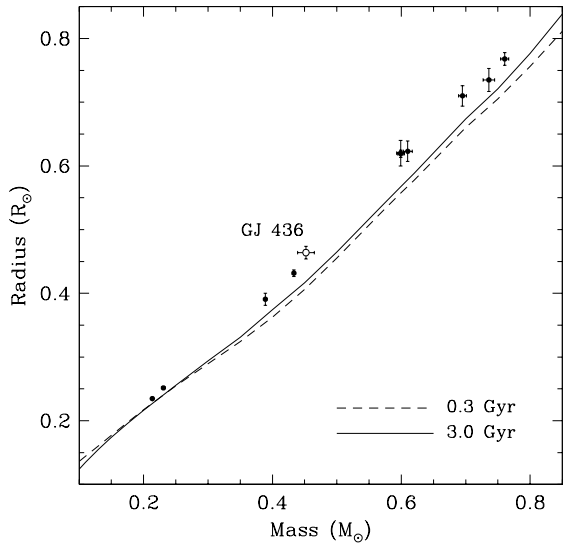}
\vskip -0.3in 

\figcaption[]{Mass-radius relation for all double-lined eclipsing
binaries with relative mass and radius errors under 3\% \citep[data
from][]{Ribas:06}. GJ~436 is shown at the values determined from our
modeling (open circle), and is seen to display the same radius
discrepancy as the other systems. Two solar-metallicity model
isochrones by \cite{Baraffe:98} are shown for reference, corresponding
to ages as labeled.\label{fig:mlr}}

\end{figure}

Our mass and radius estimates are compared with measurements for
late-type double-lined eclipsing binaries in Figure~\ref{fig:mlr}.
Only systems with the most accurate determinations are shown (relative
errors below 3\%), which are taken from the summary by
\cite{Ribas:06}. For comparison we include two isochrones from
\cite{Baraffe:98} corresponding to ages of 300 Myr (as estimated for
two of these binaries) and 3 Gyr (more representative of the
field). All eclipsing binary systems are seen to have larger sizes
than predicted for their mass.  We note also that GJ~436 lies in the
gap between masses of 0.43 M$_{\sun}$ (for CU~Cnc~A) and 0.60
M$_{\sun}$ (GU~Boo~B), and thus provides valuable additional
information on the radius discrepancies for low mass stars.

\cite{Lopez-Morales:07} has investigated how these discrepancies
($\Delta R_{\star}/R_{\star}$) depend on metallicity and the strength
of the chromospheric activity, quantified in terms of the X-ray
luminosity (specifically, $L_X/L_{\rm bol}$), for the rather limited
sample available so far. Both of these factors have been suggested to
play a role. The above study examined single M dwarfs as well as M
dwarfs in binary systems.  GJ~436 is a rather inactive star for its
type \citep{Endl:03, Butler:04}, but was detected nonetheless by ROSAT
as an X-ray source because of its proximity. The X-ray luminosity was
reported by \cite{Hunsch:99} to be $L_X = 0.7 \times 10^{27}$
erg~s$^{-1}$. When combined with the bolometric luminosity in
Table~\ref{tab:properties}, we obtain $L_X/L_{\rm bol} = 7.0 \times
10^{-6}$. The metallicity was estimated by \cite{Bonfils:05} to be
near solar: [Fe/H] $= -0.03 \pm 0.20$.  Considering GJ~436 as a single
star, the value $\Delta R_{\star}/R_{\star} \sim$ 10\% we find is
consistent with the overall conclusions of \cite{Lopez-Morales:07} in
the sense that it is similar to the offsets for other systems
regardless of $L_X/L_{\rm bol}$, and at the same time it seems to
follow the trend with [Fe/H] exhibited by other single M dwarfs.

\section{Planet parameters and final remarks}
\label{sec:discussion}

With the host star properties known, the planet parameters we infer
are given in the bottom section of Table~\ref{tab:properties}.  The
required orbital period and velocity semi-amplitude $K_{\star}$ are
adopted from \cite{Maness:07}, along with the eccentricity from
\cite{Demory:07}, who obtain a similar value for $K_{\star}$. The
improved precision of these derived parameters is a reflection of the
better stellar parameters. The slightly larger planet radius than in
previous studies confirms with even greater statistical significance
the conclusions of earlier authors regarding the presence of a
hydrogen/helium envelope \citep{Gillon:07a, Deming:07, Gillon:07b},
and agrees very well with the models by \cite{Fortney:07} for a 10\%
fraction of those elements.

In this paper we have shown that GJ~436 provides a valuable test of
stellar evolution theory near the bottom of the main sequence, made
possible by the fact that it has a transiting planet.  Traditional
studies in the area of low-mass stars have made the comparison with
models by measuring the mass and radius directly for double-lined
eclipsing systems containing M dwarfs. In a few other cases angular
diameters have been measured interferometrically for single stars, and
the mass has been inferred from empirical $M\!-\!L$ relations
\citep[e.g.,][]{Lane:01, Segransan:03}. More recently, a variety of
constraints and assumptions have been used to infer the mass and
radius of the late-type secondaries in F+M systems observed as part of
transiting planet surveys \citep[e.g.,][]{Bouchy:05, Pont:05,
Beatty:07}. Though perhaps not as compelling as having actual
model-independent mass and radius measurements, the approach in the
present work is able to make use of available information for GJ~436
and compare the models directly with the observational constraints
\emph{without} requiring a direct measurement of the mass and radius.
The discrepancy in $R_{\star}$ is derived by parameterizing it in
terms of a single adjustment factor to the model radii ($\beta$),
assuming the luminosity from theory is accurate, as other observations
seem to indicate.

NASA's upcoming \emph{Kepler\/} mission, currently slated to launch in
early 2009, will emphasize the search for transiting Earth-size
planets.  These should be easier to detect around late-type
stars. Therefore, we anticipate that many systems similar to GJ~436
could be found and become a significant source of information on radii
for low-mass stars, since they will have all the observational
constraints needed (including trigonometric parallaxes) to test models
of stellar evolution in the way we have done here.

\acknowledgements 

The anonymous referee is thanked for a prompt and helpful report. The
author acknowledges partial support for this work from NASA grant
NNG04LG89G and NSF grant AST-0708229. This research has made use of
the SIMBAD database, operated at CDS, Strasbourg, France, of NASA's
Astrophysics Data System Abstract Service, and of data products from
the Two Micron All Sky Survey, which is a joint project of the
University of Massachusetts and the Infrared Processing and Analysis
Center/California Institute of Technology, funded by NASA and the NSF.

\begin{deluxetable}{lcccc}
\tablewidth{0pc}
\tablecaption{Stellar parameters for GJ~436 based on different sets of
constraints, using the models by
\cite{Baraffe:98}.\label{tab:baraffe}}
\tablehead{  & \multicolumn{4}{c}{Observational constraints} \\ [+1.0ex]
\cline{2-5} \\ [-1.5ex]
\colhead {~~~~~~Parameter~~~~~~} & 
\colhead{$T_{\rm eff}$ and $a/R_{\star}$} &
\colhead{$J\!-\!K$ and $a/R_{\star}$} & 
\colhead{$T_{\rm eff}$ and $M_K$} &
\colhead{$J\!-\!K$ and $M_K$}}
\startdata
$M_{\star}$ (M$_{\sun}$)\dotfill & $0.497_{-0.015}^{+0.003}$    & $0.515_{-0.033}^{+0.042}$    & $0.448_{-0.008}^{+0.018}$    & $0.452_{-0.012}^{+0.014}$    \\ [+1.0ex]
$R_{\star}$ (R$_{\sun}$)\dotfill & $0.469_{-0.015}^{+0.000}$    & $0.484_{-0.030}^{+0.035}$    & $0.421_{-0.009}^{+0.009}$    & $0.421_{-0.010}^{+0.009}$    \\ [+1.0ex]
$\log g_{\star}$ (cgs)\dotfill   & $4.792_{-0.000}^{+0.020}$    & $4.780_{-0.032}^{+0.032}$    & $4.841_{-0.009}^{+0.020}$    & $4.843_{-0.011}^{+0.018}$    \\ [+1.0ex]
$a/R_{\star}$\dotfill            & \nodata                      & \nodata                      & $14.63_{-0.19}^{+0.34}$      & $14.65_{-0.21}^{+0.31}$      \\ [+1.0ex]
$T_{\rm eff}$ (K)\dotfill        & \nodata                      & $3684_{-55}^{+87}$           & \nodata                      & $3585_{-13}^{+19}$           \\ [+1.0ex]
$J\!-\!K$ (mag)\tablenotemark{a}\dotfill         & $0.801_{-0.001}^{+0.008}$    & \nodata                      & $0.810_{-0.007}^{+0.002}$    & \nodata                      \\ [+1.0ex]
$M_V$ (mag)\dotfill              & $9.86_{-0.00}^{+0.12}$    & $9.71_{-0.36}^{+0.27}$    & $10.256_{-0.094}^{+0.070}$   & $10.244_{-0.082}^{+0.082}$   \\ [+1.0ex]
$M_K$ (mag)\tablenotemark{a}\dotfill              & $5.758_{-0.001}^{+0.090}$    & $5.66_{-0.22}^{+0.18}$    & \nodata                      & \nodata                      \\ [+1.0ex]
$L_{\star}$ (L$_{\sun}$)\dotfill & $0.0348_{-0.0035}^{+0.0000}$ & $0.0383_{-0.0069}^{+0.0099}$ & $0.0260_{-0.0016}^{+0.0015}$ & $0.0260_{-0.0017}^{+0.0014}$ \\ [-1.0ex]
\enddata
\tablenotetext{a}{Magnitudes are in the CIT photometric system of \cite{Elias:82}.}
\end{deluxetable}

\begin{deluxetable}{lc}
\tablewidth{0pc}
\tablecaption{Stellar and planetary parameters for the GJ~436 system.\label{tab:properties}}
\tablehead{\colhead{~~~~~~~~~~~~~~Parameter~~~~~~~~~~~~~~} & \colhead{Value}}
\startdata
\noalign{\vskip -5pt}
\sidehead{Stellar parameters}
~~~$M_{\star}$ (M$_{\sun}$)\dotfill & $0.452_{-0.012}^{+0.014}$ \\ [+1.0ex]
~~~$R_{\star}$ (R$_{\sun}$)\dotfill & $0.464_{-0.011}^{+0.009}$ \\ [+1.0ex]
~~~$L_{\star}$ (L$_{\sun}$)\dotfill & $0.0260_{-0.0017}^{+0.0014}$ \\ [+1.0ex]
~~~$\log g_{\star}$ (cgs)\dotfill   & $4.843_{-0.011}^{+0.018}$ \\ [+1.0ex]
~~~Age (Gyr)\tablenotemark{a}\dotfill                & $6_{-5}^{+4}$ \\
\noalign{\vskip -2pt}
\sidehead{Planetary parameters}
~~~$M_p$ (M$_{\earth}$)\dotfill & 23.17~$\pm$~0.79\phn \\ [+0.7ex]
~~~$R_p$ (R$_{\earth}$)\dotfill & $4.22_{-0.10}^{+0.09}$ \\ [+1.0ex]
~~~$\rho_p$ (g~cm$^{-3}$)\dotfill  & $1.69_{-0.12}^{+0.14}$ \\ [+0.7ex]
~~~$\log g_p$ (cgs)\dotfill        & 3.107~$\pm$~0.040 \\ [+0.7ex]
~~~$a$ (AU)\dotfill              & 0.02872~$\pm$~0.00027 \\ [-1.5ex]
\enddata
\tablenotetext{a}{Due to the unevolved nature of GJ~436 the age is
essentially unconstrained by the observations. We list this value
only for completeness.}
\end{deluxetable}

\end{document}